\documentclass[prl, twocolumn,superscriptaddress,nobibnotes]{revtex4}
\usepackage{graphicx}
\usepackage{amssymb}
\usepackage{amsmath}
\usepackage{bm}
\usepackage{color}
\usepackage{float}
\usepackage{setspace}
\usepackage{physics}
\usepackage{subfigure}
\usepackage[utf8]{inputenc}   
\usepackage[T1]{fontenc}

\large

\begin{document}

\title{Observation of anomalous exciton polariton bands in \\ PEPI perovskite based microcavity at room temperature}

\author{Chunzi Xing}
\affiliation{Department of Physics, School of Science, Tianjin University, Tianjin 300072, China}

\author{Xiaokun Zhai}
\affiliation{Department of Physics, School of Science, Tianjin University, Tianjin 300072, China} 

\author{Chenxi Yang}
\affiliation{Department of Physics, School of Science, Tianjin University, Tianjin 300072, China}

\author{Peilin Wang}
\affiliation{Center for Applied Mathematics and KL-AAGDM, Tianjin University, Tianjin, 300072, China} 

\author{Jiaxiang Mu}
\affiliation{Department of Physics, School of Science, Tianjin University, Tianjin 300072, China}

\author{Xinmiao Yang}
\affiliation{Department of Physics, School of Science, Tianjin University, Tianjin 300072, China} 

\author{Yao Li}
\affiliation{Department of Physics, School of Science, Tianjin University, Tianjin 300072, China}

\author{Xianxiong He}
\affiliation{College of New Materials and Chemical Engineering, Beijing Institute of Petrochemical Technology, Beijing 102617, China.  }

\author{Yong Zhang}
\affiliation{Center for Applied Mathematics and KL-AAGDM, Tianjin University, Tianjin, 300072, China} 

\author{Haitao Dai}
\affiliation{Department of Physics, School of Science, Tianjin University, Tianjin 300072, China}

\author{Liefeng Feng}
\affiliation{Department of Physics, School of Science, Tianjin University, Tianjin 300072, China}

\author{Tingge Gao}
\affiliation{Department of Physics, School of Science, Tianjin University, Tianjin 300072, China}

\begin{abstract}
Recently anomalous energy bands with negative mass attract intensive attention where non-Hermiticity plays an important role. In this work we observe anomalous exciton polariton bands in PEPI perovskite based microcavity at room temperature. We simulate the anomalous band structure using a non-Hermitian coupled oscillator model which agree with experiments very well. Our results offer to study non-Hermitian polariton wave dynamics at room temperature.
\end{abstract}

\maketitle

\section*{1. Introduction}


Non-Hermiticity is intensively investigated during last decade in plenty of different physical systems, including optics, electrons, cold atoms, acoustics and hybrid light matter systems. Non-Hermiticity can inevitably affect these systems' modal profile and energy landscape. For example, novel photon dynamics in \textit{PT} symmetric synthetic lattices \cite{4PT synthetic lattice} or microcavities \cite{5PT yanglan}, single mode \cite{6single mode laser xiangzhang, 7single mode laser CDN} and vortex lasing \cite{8PT vortex laser}, unidirectional propagation in a waveguide \cite{9unidirectional PT, 41}, suppression and revival of lasing in coupled whispering-gallery-mode microcavities \cite{10loss revival  yanglan}, optical isolation \cite{13PT-xiaomin}, and an exceptional ring \cite{14exceptional circle}. In addition, non-Hermiticity can modify the energy mode structure intensively. For example, anomalous band structure with negative mass can be observed in optical microcavity where exciton polariton can be formed, which is a quasiparticle due to the strong coupling between the exciton and photon mode. In Fabry-Perot microcavity, exciton polaritons can demonstrate similar Bose-Einstein condensate process as the cold boson atoms at much higher temperature due to the very light effective mass \cite{polariton BEC1, polariton BEC2}. With dissipative coupling, the polariton dispersion can show level attraction behavior, which leads to opposite particle propagation compared with normal microcavities. Although this kind of anomalous dispersion \cite{anomalous dispersionmodel} has been observed in TMD monolayer based microcavity \cite{anomalous dispersionTMD} and GaAs microcavities \cite{anomalous dispersionGaAs}, it has not been reported in perovskite microcavity yet. This gap motivates the investigation of perovskite microcavities, which offer unique advantages for room-temperature non-Hermitian photonics.

In the study of non-Hermitian physics, the choice of semiconductor material is crucial for achieving efficient light-matter coupling. Traditional III-V semiconductors such as CdTe \cite{polariton BEC1} and GaAs \cite{GaAs-GaAlAs} offer excellent exciton bonding energy and controllable cavity structures, but their excitonic stability is typically limited to cryogenic temperatures. Organic semiconductors exhibit good excitonic luminescence efficiency and chemical tunability  \cite{organic semiconductor,organic semiconductor microcavity}, yet they often suffer from poor thermal and environmental stability. Transition metal dichalcogenides (TMDs) demonstrate strong excitonic effects at room temperature due to their layered structure, but the fabrication of large-area homogeneous films remains challenging. All-inorganic perovskites such as  CsPbBr$_3$ show high luminescence efficiency and stability\cite{gao ying, Xiaokun vortex}, yet their relatively low exciton binding energy and weak optical anisotropy limit polarization-dependent tunability in microcavities.

In contrast, two-dimensional Ruddlesden–Popper perovskites (RPPVs) combine the advantages of organic and inorganic components, exhibiting not only high exciton binding energy—supporting stable exciton polariton formation at room temperature—but also excellent photostability, ease of low-cost large-area fabrication, and flexible bandgap tunability. Among them, fluorinated (PEA)$_2$PbI$_4$ (PEPI) demonstrates good photochemical stability and exhibits excellent strong coupling properties in microcavity photonics. The materials have shown broad application potential in photovoltaics, photodetection, light-emitting diodes, and beyond. Their introduction into microcavity photonics offers new possibilities for realizing room-temperature polaritonic devices. Particularly in planar microcavity structures, the high refractive index and strong exciton–photon coupling capability of PEPI make it an ideal optical medium, supporting high-quality optical modes and enabling flexible tuning of multimode polariton dispersion through its layered structure\cite{pepi1,pepi2,peai3}. This provides a material foundation for designing photonic structures with non-Hermitian features, such as exceptional points and nonreciprocal transmission.

In this work, we report an optical microcavity system based on the layered perovskite material (PEPI), with its structural configuration illustrated in Figure 1(a). By constructing this cavity structure, we observe anomalous exciton-polariton bands at room temperature, as schematically represented in Figure 1(b). We further developed a non-Hermitian coupled oscillator model, which successfully reproduces the experimentally observed anomalous band structure and shows excellent agreement with the measurements. This study provides a new platform for investigating non-Hermitian polariton wave dynamics at room temperature and offers new perspectives for designing tunable non-Hermitian photonic devices.

\begin{figure}[t]
		\centering	\includegraphics[width=0.8\linewidth]{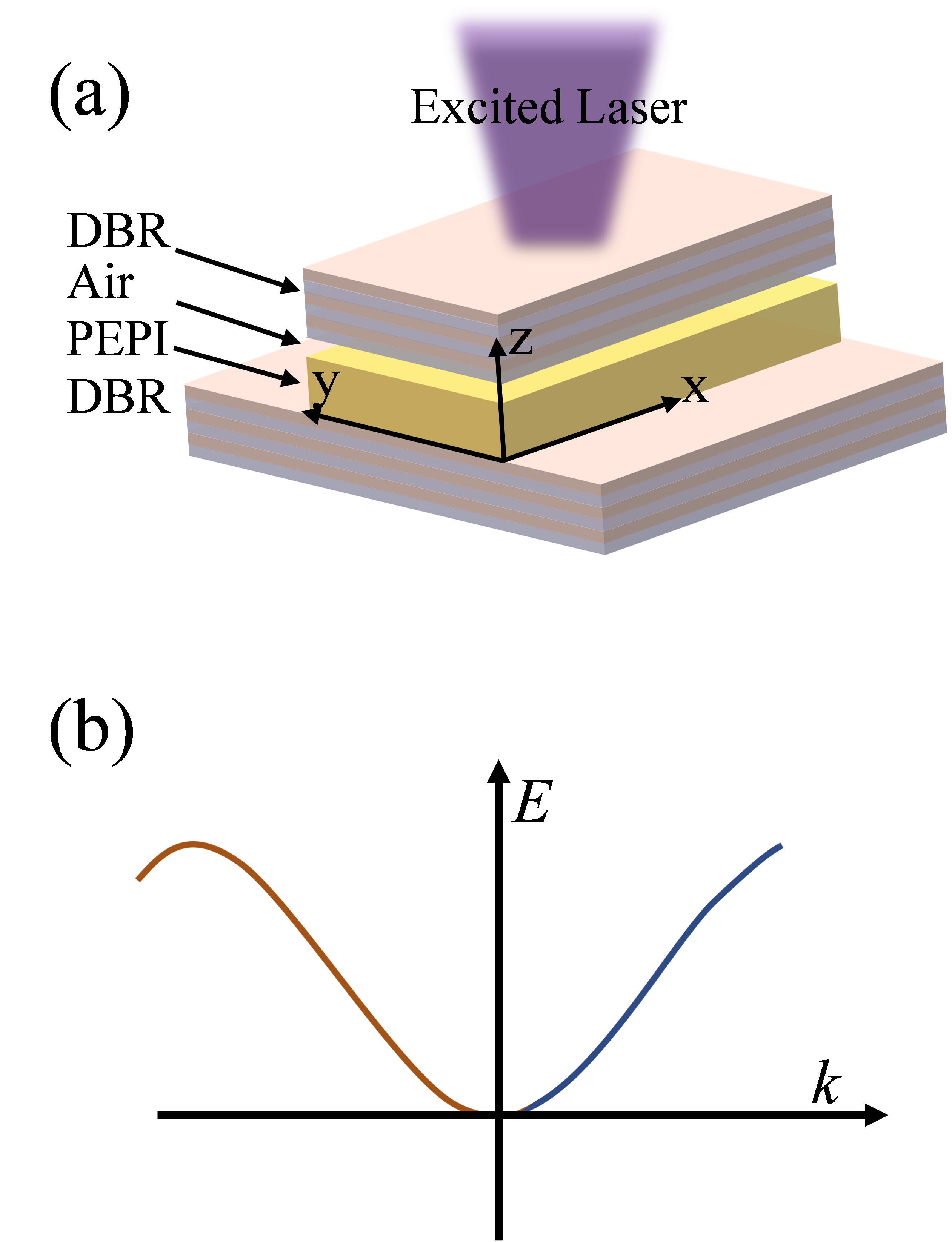}
	\caption{\textbf{{Schematic graph of the PEPI perovskite microcavity.}} (a) The structure of the PEPI microcavity. (b) Exciton polariton band structure, exhibiting conventional dispersion along the k direction or anomalous dispersion along the k direction, depending on the non-Hermiticity of the microcavity.} 
\end{figure}


\section{2. R\lowercase{esults and} D\lowercase{iscussion}}

To achieve clear observation of polaritons in a PEPI perovskite microcavity, this work employs high-quality single crystals rather than polycrystalline thin films, ensuring the formation of complete crystalline facets with smooth interfaces. The two-dimensional organic–inorganic hybrid perovskite material used in the experiment is (PEA)$_2$PbI$_4$ (abbreviated as PEPI), which adopts a typical Ruddlesden–Popper layered perovskite structure. In this material, phenethylammonium cations (PEA$^{+}$, C$_6$H$_5$CH$_2$CH$_2$NH$_3$$^{+}$) serve as organic spacer layers inserted between inorganic [PbI$_6$]$^{4-}$ octahedral sheets, forming a stable two-dimensional quantum-well architecture. PEPI single crystals are synthesized directly on a distributed Bragg reflector (DBR) substrate using a solution-based method \cite{growth1,growth2,growth3}, which facilitates the growth of large-area, high-crystallinity layered single-crystalline films and provides a solid foundation for subsequent microcavity construction and optical characterization. 

\begin{figure}[t]
		\centering	\includegraphics[width=0.99\linewidth]{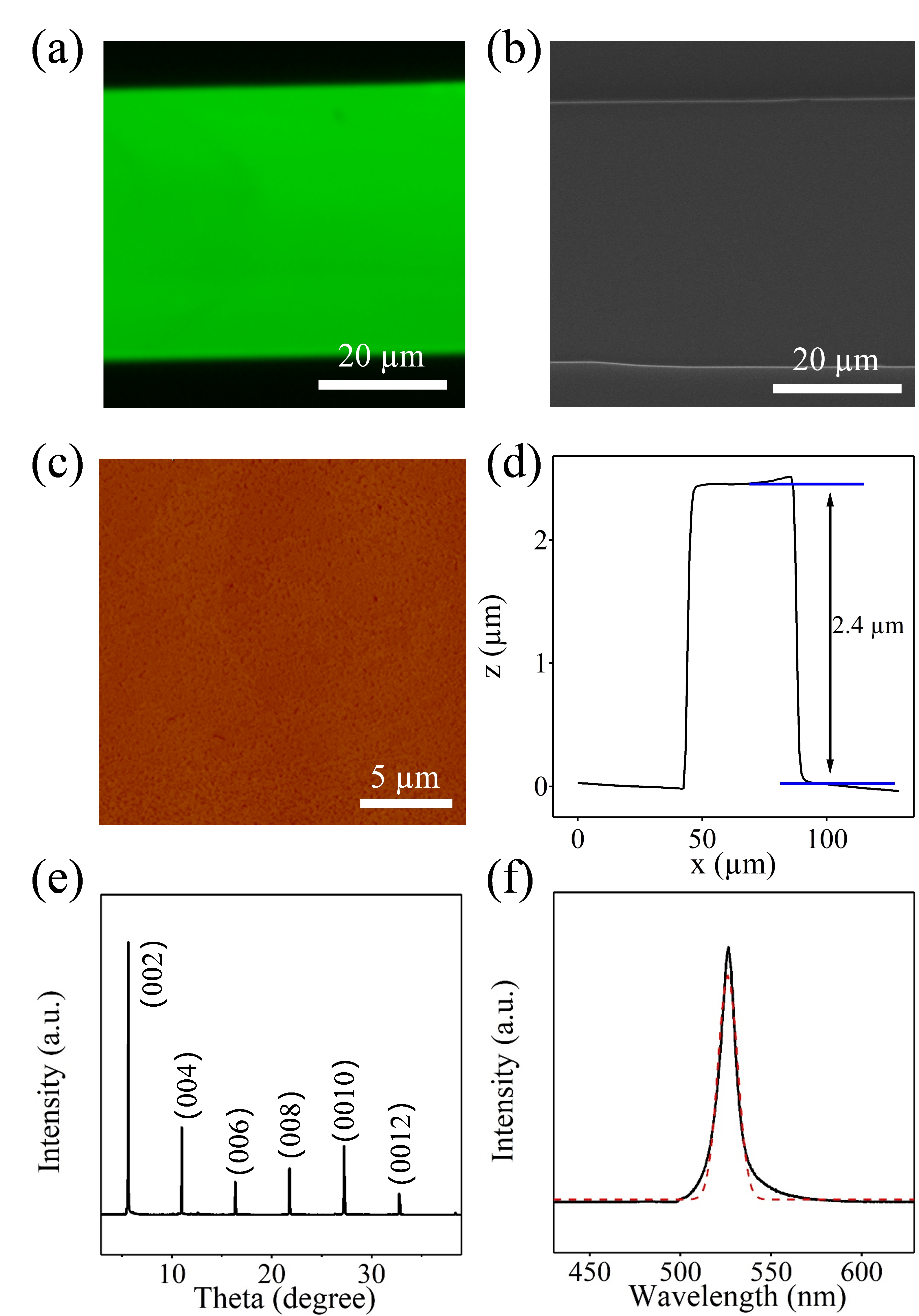}
	\caption{\textbf{{Structural and optical characterization of PEPI perovskite crystals grown on a DBR substrate.}} (a) Fluorescence microscopy image of the as-grown PEPI crystals. (b) Corresponding SEM image showing the platelet-like morphology.(c) AFM topography of the crystal surface. (d) Height profile across the step edge, revealing the thickness of the PEPI. (e) XRD pattern of the PEPI perovskite film, confirming its single-crystal characteristics. (f) Room-temperature photoluminescence (PL) spectrum of PEPI, with the red dashed curve indicating the Gaussian fit.} 
\end{figure}

Systematic morphological and structural characterizations are conducted to verify the quality of the PEPI synthesized on bottom DBR. Figure 2(a) presents a fluorescence microscopy image of the sample, showing uniform crystal coverage. Scanning electron microscopy (SEM) images (Figure 2(b)) reveal well-defined platelet-like structures with smooth surfaces and sharp boundaries. Atomic force microscopy (AFM) measurements (Figure 2(c)) further confirm the surface flatness, with height-profile line scans indicating a uniform PEPI thickness of approximately 2.4 $\mu$m, as shown in Figure 2(d). Additionally, X-ray diffraction (XRD) patterns (Figure 2(e)) display sharp diffraction peaks, confirming the single-crystalline nature of the PEPI perovskite. As presented in Figure 2(f), photoluminescence (PL) spectroscopy exhibits a narrow emission peak centered at 526 nm with a width at half maximum (FWHM) of 15 nm. These characterization results collectively demonstrate that the synthesized PEPI single crystals possess excellent crystallinity, making them suitable for subsequent microcavity fabrication and optical studies.

\begin{figure}[t]
		\centering	\includegraphics[width=0.9\linewidth]{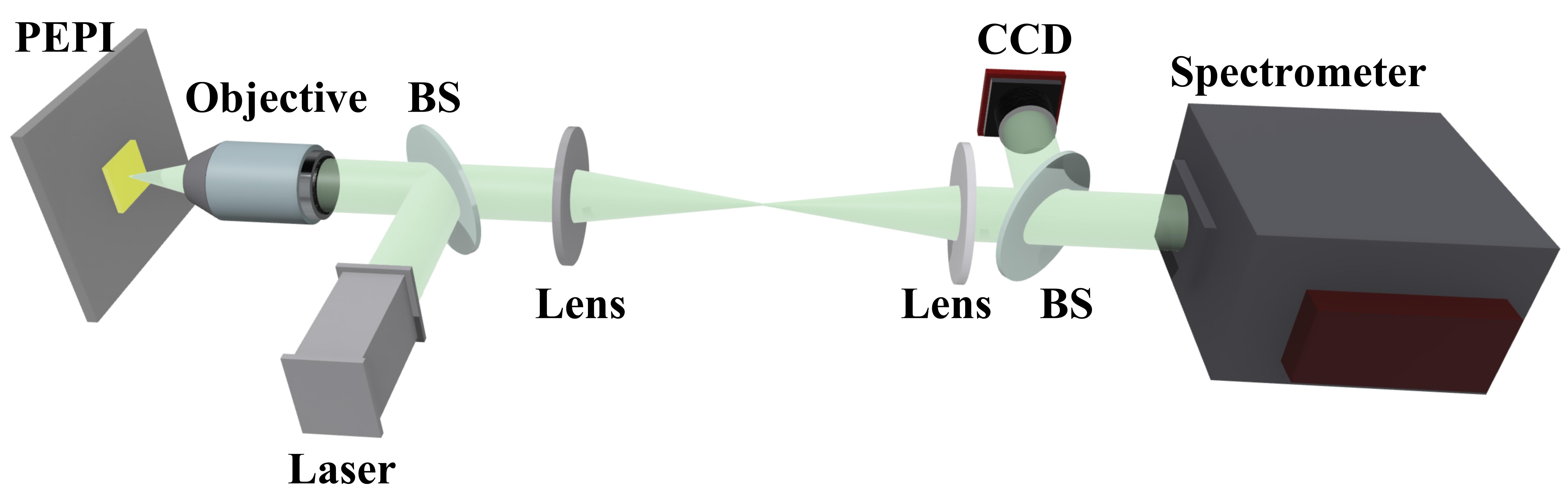}
	\caption{\textbf{Schematic of the home-built ARPL spectra system.}} 
\end{figure}

In the experiment, to systematically investigate the optical properties of PEPI perovskite crystals grown on the bottom DBR and their polaritonic behavior in a microcavity structure, we construct an angle-resolved spectroscopy measurement setup based on a dual-lens imaging system (Figure 3). The optical system employs a reflective geometry, enabling momentum-space spectral acquisition from normal incidence to large angles. Additionally, a continuous-wave laser (410 nm) is collimated and directed onto the sample surface. By precisely rotating the sample stage or adjusting the incident beam angle, spectra at different in-plane wavevectors \( k_{\parallel} \) are probed. The emitted light signal is collected via a lens and coupled into an imaging spectrometer, ultimately obtaining angle-dependent photoluminescence (PL) or reflectance spectra.

Figure 4 presents the dispersion characteristics of exciton polaritons in the PEPI perovskite microcavity along the k$_x$ (Figure 4(a)) and wavevector directions, where the energy-wavevector (E-k) spectral distribution is characterized by angle-resolved photoluminescence (ARPL) spectroscopy (Figure 3), and the color mapping of intensity intuitively reflects the intensity evolution of exciton-polariton modes. multiple distinct lower polariton branches are observed in the spectra, each corresponding to strong coupling between different photonic modes and excitons. The spectral splitting in the low-wavevector region (|k| < 4 $\mu$m$^{-1}$) originates from the in-plane anisotropy of the PEPI perovskite crystal, leading to the differentiation of exciton-photon coupling modes. In the high-wavevector region (|k| > 4 $\mu$m$^{-1}$), the dispersion exhibit two distinct curvature characteristics: the negative curvature branch corresponds to a negative effective mass, reflecting dissipative coupling due to the non-Hermiticity within the microcavity; the positive curvature branch corresponds to a positive effective mass, which is a typical dispersion behavior of exciton polaritons near the conduction band minimum. With more positive detuning with more exciton components, the two branches show anomalous dispersion with negative mass together. 

\begin{figure}[t]
		\centering	\includegraphics[width=0.9\linewidth]{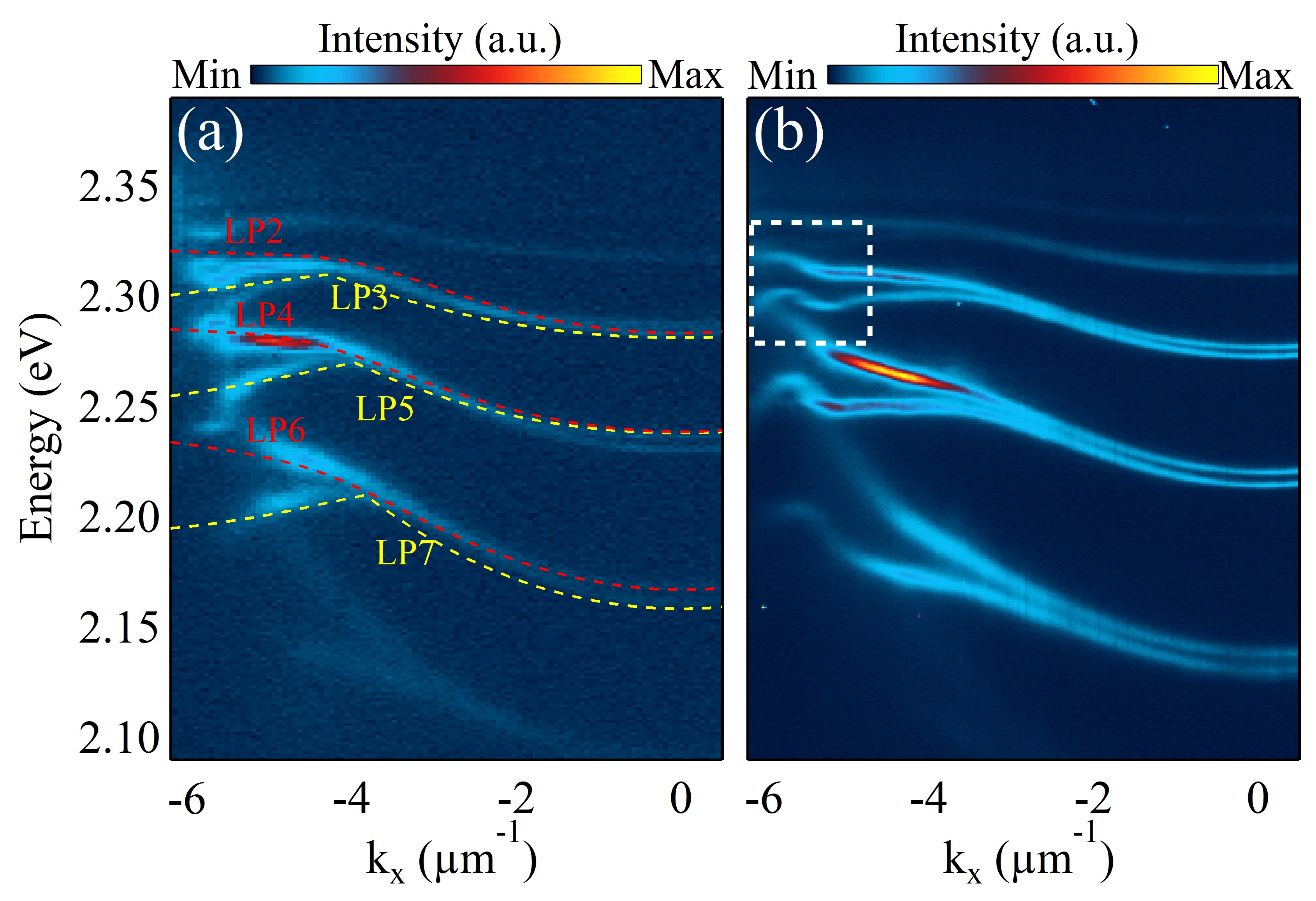}
	\caption{\textbf{{Anomalous exciton polariton bands of the PEPI-based microcavity.}} (a) Angle-resolved photoluminescence (ARPL) dispersion spectra of exciton polaritons in PEPI perovskite. The dashed lines are the fitted dispersion branches of exciton polariton modes. (b) Anomalous dispersion curves observed along the k$_x$ direction at a different sample position. The white dashed box highlights regions where the sign of the effective mass changes twice. The color bar indicates the photoluminescence intensity.} 
\end{figure}

To simulate the dipersions in the experiment, we use a non-Hermitian coupled oscillator model. The system dynamic in momentum k-space is described by the non-Hermitian Hamiltonian:
\begin{equation}
H(\mathbf{k}) =
\begin{pmatrix}
 E_{\mathrm{cav}}(\mathbf{k}) + \mathrm{i}\Gamma_1 & \Omega +  \mathrm{i}\Gamma_3 \\
 \Omega -  \mathrm{i}\Gamma_3 & E_{\mathrm{exc}} +  \mathrm{i}\Gamma_2
\end{pmatrix},
\end{equation}
where $E_{\mathrm{cav}}(\mathbf{k}) = \hbar^2\mathbf{k}^2/(2m_{\mathrm{eff}}) + \delta$, with $\mathbf{k}^2 = k_x^2 + k_y^2$, describes the cavity photon dispersion. Here, $m_{\mathrm{eff}}$ is the effective photon mass arising from cavity confinement, and $\delta$ denotes the cavity-exciton detuning at zero momentum. The exciton energy $E_{\mathrm{exc}}$ and the Rabi coupling strength $\Omega$ are momentum-independent constants. The imaginary terms $\Gamma_1$, $\Gamma_2$ and $\Gamma_3$ are constant decay rates characterizing the cavity mode, exciton mode, and their coupling, respectively. Notably, $\Gamma_1$ plays a crucial role in determining the dispersion characteristics: smaller values preserve the smooth parabolic form of the bare cavity dispersion, while larger values introduce significant non-Hermitian effects that lead to dispersion bending and anomalous curvature.

To compute the polariton dispersion, we diagonalize the Hamiltonian numerically in momentum space. All spatial derivatives are evaluated via Fourier multipliers, allowing $H(\mathbf{k})$ to act pointwise without assembling large matrices. We set $\hbar = 1.0546 \times 10^{-34}$~J$\cdot$s and adopt experimentally relevant parameters: $m_{\mathrm{eff}} = 1.6 \times 10^{-5} m_e$ (where $m_e$ is the free electron mass), $\Omega = 0.08$~meV, $\Gamma_2 = 0.01$~meV, and $\Gamma_3 = 0.01$~meV. The cavity decay rate $\Gamma_1$ is treated as a tunable parameter to capture different physical regimes.

We perform the simulations on a rectangular spatial domain:
\begin{equation}
D = [-L_x/2, L_x/2] \times [-L_y/2, L_y/2]
\end{equation}
with $L_x = L_y = 100~\mu\mathrm{m}$, discretized uniformly on $N_x = N_y = 256$ grid points. The corresponding momentum-space grid is constructed via the discrete Fourier transform, with components:
\begin{align*}
k_x &= \frac{2\pi}{L_x}\left[-\frac{N_x}{2}, -\frac{N_x}{2}+1, \ldots, \frac{N_x}{2}-1\right], \\
k_y &= \frac{2\pi}{L_y}\left[-\frac{N_y}{2}, -\frac{N_y}{2}+1, \ldots, \frac{N_y}{2}-1\right],
\end{align*}
where the FFT-shift operation is applied to center the zero-frequency component. For the dispersion analysis, we extract two one-dimensional slices at $k_x = 0$ and $k_y = 0$, respectively.

At each $k_y$, diagonalization of the $2 \times 2$ Hamiltonian yields two complex eigenvalues, which we classify as lower polariton (LP) and upper polariton (UP) branches according to the real parts of their energies. The dispersion characteristics depend strongly on the cavity decay rate $\Gamma_1$, which controls the degree of non-Hermiticity.

To reproduce experimentally observed dispersion features, we employ different values of $\Gamma_1$ depending on the regime of interest. For smooth, parabolic dispersion curves with minimal distortion in high-quality cavities, we set $\Gamma_1 = 0.15$~meV, corresponding to the low-loss regime where non-Hermitian effects are weak. Conversely, to capture pronounced dispersion bending and anomalous curvature in systems with significant dissipation, we use $\Gamma_1 = 0.2$~meV. This variation reflects the experimental reality that cavity quality factors vary substantially with fabrication conditions and material properties, leading to qualitatively different dispersion behaviors.


Finally, we note that more anomalous dispersions with changing the sign of the effective mass twice at large wavevector region along $k_x$ direction can be observed in other area of the microcavity due to thickness variance of the PEPI microplate, as plotted in Figure 4(b).

\section{3. C\lowercase{onclusion}}

In summary, we observe the non-Hermiticity induced anomalous exciton polariton dispersion in a PEPI based microcavity. The anomalous dispersion induces negative mass at large wavevector region. In addition, the dispersion can show more complex negative mass behavior due to thickness variation of PEPI perovskites. Our results reveal non-Hermitian dispersions in perovskite based microcavity.

\section*{Acknowledgments}
\begin{acknowledgments}
T. Gao acknowledges the National Natural Science Foundation of China (NSFC, No. $12174285$, $12474315$). X. Zhai acknowledges the support from the National Natural Science Foundation of China(NSFC, No. 12504372) and the China Postdoctoral Science Foundation-Tianjin Joint Support Program under Grant Number (No. 2025T003TJ). 
\end{acknowledgments}


\begin{thebibliography}{99}


\bibitem{4PT synthetic lattice}A. Regensburger, C. Bersch, M. A. Miri, G. Onishchukov, D. N. Christodoulides and U. Peschel, Parity–time synthetic photonic lattices. \textit{Nature}  \textbf{488}, 167-171 (2012).

\bibitem{5PT yanglan} B Peng, \c{S}. K. {\"O}zdemir, F. Lei, F. Monifi, M. Gianfreda, G. L. Long, S. Fan, F. Nori, C. M. Bender and L. Yang, Parity–time-symmetric whispering-gallery microcavities. \textit{Nature Physics} \textbf{10}, 94-398 (2014).

\bibitem{6single mode laser xiangzhang} L. Feng, Z.J. Wong, R.M. Ma, Y. Wang and X. Zhang, Single-mode laser by parity-time symmetry breaking. \textit{Science} \textbf{346}, 972-975 (2014).

\bibitem{7single mode laser CDN} H. Hodaei, M. A.  Miri, M. Heinrich, D. N. Christodoulides and M. Khajavikhan, Parity-time–symmetric microring lasers. \textit{Science} \textbf{346}, 975-978 (2014).

\bibitem{8PT vortex laser} P. Miao, Z. Zhang, J. Sun, W. Walasik, S. Longhi, N. M. Litchinitser and L. Feng, Orbital angular momentum microlaser. \textit{Science} \textbf{353}, 464-467 (2016).

\bibitem{9unidirectional PT} L. Feng, Y. L  Xu, W, S. Fegadolli, M. H. Lu, J. E. Oliveira, V. R. Almeida, Y. Chen and A. Scherer, Experimental demonstration of a unidirectional reflectionless parity-time metamaterial at optical frequencies. \textit{Nature Materials} \textbf{12}, 108-113 (2013).

\bibitem{41} L. Jin and Z. Song, Incident direction independent wave propagation and unidirectional lasing. \textit{Physical Review Letters} \textbf{121}, 073901 (2018).

\bibitem{10loss revival  yanglan} B Peng, \c{S}. K. {\"O}zdemir, S. Rotter, H. Yilmaz, M. Liertzer, F. Monifi, C. Bender, F. Nori and L. Yang, Loss-induced suppression and revival of lasing. \textit{Science} \textbf{346}, 328-332 (2014).


\bibitem{13PT-xiaomin} L. Chang, X. Jiang, S. Hua, C. Yang, J. Wen, L. Jiang, G. Li, G. Wang and M. Xiao, Parity–time symmetry and variable optical isolation in active–passive-coupled microresonators. \textit{Nature Photonics} \textbf{8}, 524-529 (2014).

\bibitem{14exceptional circle} B. Zhen, C. W. Hsu, Y. Igarashi, L. Lu, I. Kaminer, A. Pick, S. L. Chua, J. D. Joannopoulosand M. Solja{\v c}i{\' c}, Spawning rings of exceptional points out of Dirac cones. \textit{Nature} \textbf{525}, 354-358 (2015).


\bibitem{polariton BEC1} J. Kasprzak, M. Richard, S. Kundermann, A. Baas, P. Jeambrun, J. M. J. Keeling, F. M. Marchetti, M. H. Szyma{\' n}ska, R. Andr{\'e}, J. L. Staehli, V. Savona, P. B. Littlewood, B. Deveaud and  L. S. Dang, Bose–Einstein condensation of exciton polaritons. \textit{Nature} \textbf{443,} 409-414 (2006).

\bibitem{polariton BEC2} R. Balili, V. Hartwell, D. Snoke, L. Pfeiffer and K. West, Bose-Einstein condensation of microcavity polaritons in a trap. \textit{Science} \textbf{316,} 1007-1010 (2007).

\bibitem{anomalous dispersionmodel}O. Bleu, K. Choo, J. Levinsen, M. M. Parish, Dissipative light-matter coupling and anomalous dispersion in nonideal cavities. \textit{Physical Review A} \textbf{109,} (2024).

\bibitem{anomalous dispersionTMD} M. Wurdack, T. Yun, M. Katzer, A. G. Truscott, A. Knorr, M. Selig, E. A. Ostrovskaya, E. Estrecho, Negative-mass exciton polaritons induced by dissipative light-matter coupling in an atomically thin semiconductor. \textit{Nature Communications} \textbf{14,} (2023). 


\bibitem{anomalous dispersionGaAs} D. Biegańska, M. Pieczarka, C. Schneider, S. Höfling, S. Klembt, and M. Syperek, Anomalous dispersion via dissipative coupling in a quantum well exciton-polariton microcavity. arXiv:2404.14116v2 [cond-mat.mes-hall] 19 Sep 2024. 





\bibitem{GaAs-GaAlAs} D. Bajoni, P. Senellart, E. Wertz, I. Sagnes, A. Miard, A. Lemaitre and J. Bloch, Polariton laser using single micropillar GaAs-GaAlAs semiconductor cavities. \textit{Physical Review Letters} \textbf{100,} 047401 (2008).
 
 
\bibitem{organic semiconductor} D. G. Lidzey, D. D. Bradley, A. Armitage, S. Walker, M. S. Skolnick, Photon-mediated hybridization of frenkel excitons in organic semiconductor microcavities. \textit{Science} \textbf{288,} 1620-1623 (2000).

\bibitem{organic semiconductor microcavity} D. G. Lidzey, D. D. C. Bradley, M. S. Skolnick, T. Virgili, S. Walker, D. M. Whittaker, Strong exciton–photon coupling in an organic semiconductor microcavity. \textit{Nature} \textbf{398,} 53-55 (1998).


\bibitem{gao ying} Y. Gao, X. Ma, X. Zhai, C. Xing, M. Gao, H. Dai, H. Wu, T. Liu, Y. Ren, X. Wang, A. Pan, W. Hu, S. Schumacher, and T. Gao, Single-shot spatial instability and electric control of polariton condensates at room temperature. \textit{Physical Review B}  \textbf{108,} 205303 (2023).


\bibitem{Xiaokun vortex} X. Zhai, X. Ma, Y. Gao, C. Xing, M. Gao, H. Dai, X. Wang, A. Pan, S. Schumacher, T. Gao, Electrically controlling vortices in a neutral exciton-polariton condensate at room temperature. \textit{Physical Review Letters} \textbf{131,} 136901  (2023).

\bibitem{pepi1} M. K\c{e}dziora, M. Kr{\' o}l, P. Kapu{\' s}ci{\' n}ski, H. Sigur\dh{}sson, R. Mazur, W. Piecek, J. Szczytko, M. Matuszewski, A. Opala, B. Pi\c{e}tka, Non-Hermitian polariton–photon coupling in a perovskite open microcavity.  \textit{Nanophotonics} \textbf{13,} 2491-2500 (2024).

\bibitem{pepi2} R. Fei, M. P. Hautzinger, A. H. Rose, Y. Dong, I. I. Smalyukh, M. C. Beard, J. van de Lagemaat, Controlling Exciton/Exciton Recombination in 2-D Perovskite Using Exciton–Polariton Coupling. \textit{The Journal of Physical Chemistry Letters} \textbf{15,} 1748-1754 (2024).

\bibitem{peai3}	 L. Polimeno, A. Fieramosca, G. Lerario, M. Cinquino, M. De Giorgi, D. Ballarini, F. Todisco, L. Dominici, V. Ardizzone, M. Pugliese, C. T. Prontera, V. Maiorano, G. Gigli, L. De Marco, D. Sanvitto,Observation of Two Thresholds Leading to Polariton Condensation in 2D Hybrid Perovskites. \textit{Advanced Optical Materials} \textbf{8,} (2020).

\bibitem{growth1}	Y. Wei, J. S. Lauret, L. Galmiche, P. Audebert, E. Deleporte, Strong exciton-photon coupling in microcavities containing new fluorophenethylamine based perovskite compounds. \textit{Optics Express} \textbf{20,} (2012).

\bibitem{growth2}	F. Lédée, G. Trippé-Allard, H. Diab, P. Audebert, D. Garrot, J.-S. Lauret, E. Deleporte, Fast growth of monocrystalline thin films of 2D layered hybrid perovskite.  \textit{CrystEngComm} \textbf{19,} 2598-2602  (2017).

\bibitem{growth3}	E. Ren, C. Zhang, F. Wang, J. Kong, L. Li, J. Chen, J. Xu, Y. zhang, Synthesis of Cs$^{+}$-Tuned (PEA)$_2$PbI$_4$ Perovskite Thin Films by One-Step Spin Coating. \textit{ECS Journal of Solid State Science and Technology} \textbf{12,} (2023).

\end{thebibliography}
\end{document}